
\vsize=7.5in
\hsize=6.6in
\hfuzz=20pt
\tolerance 10000

\baselineskip 12pt plus 1pt minus 1pt
\pageno=0

\def\){]}
\def\({[}
\def\sno{\smallskip\noindent}

\centerline{\bf Application of M\"ossbauer-Type Sum Rules for $B$ Meson
Decays}
\author{Harry J. Lipkin}
\sno
\centerline{Department of Nuclear Physics}
\centerline{\it Weizmann Institute of Science}
\centerline{Rehovot 76100, Israel}

\centerline{and}

\centerline{School of Physics and Astronomy}
\centerline{Raymond and Beverly Sackler Faculty of Exact Sciences
}
\centerline{\it Tel Aviv University}
\centerline{Tel Aviv, Israel}

\centerline{January 26, 1993}
\vskip 0.2in

\abstract

\medskip

Sum rules originally derived for the M\"ossbauer Effect are applied
to weak semileptonic B decays. The sum rules follow from assuming that
the decay by electroweak boson emission of an
unstable nucleus or heavy quark in a bound system is
described by a pointlike coupling to a
current which acts only on the decaying object, that the
Hamiltonian of the bound state depends on the momentum of the decaying
object only in the kinetic energy and that the boson has no final state
interactions. The decay rate and the first and second moments of the
boson energy spectrum for fixed momentum transfer are shown to be the
same as for a noninteracting gas of such unstable objects with a momentum
distribution the same as that of the bound state. B meson semileptonic
decays are shown to be dominated by the lowest-lying states in
the charmed meson spectrum.
\endpage

Simple features of the spectroscopy and weak transitions of hadron states
containing heavy quarks have recently been discussed in detail
\REFS{\WISGUR}{Nathan Isgur and Mark B. Wise, Phys. Lett. B 232 (1989)
113; B237 (1990) 527} \REFSCON{\WISPECT}{Nathan Isgur and Mark B. Wise,
Phys. Rev. Lett. 66 (1991) 1130}[{\WISGUR},{\WISPECT}]. The contributions
to a Bjorken sum rule
\REF{\BJSumR }{J. D. Bjorken, in Proceedings of the Rencontre de Physique de
la Vallee D'Aoste, La Thuile, Italy, [SLAC Report No. SLAC-PUB-5278, 1990 (to
be published)]}[{\BJSumR}] have been examined\REFS{\WISBAR}{Nathan Isgur, Mark
B. Wise and Michael Youssefmir, Phys. Lett. B 254 (1991) 215}
\REFSCON{\WISBJ}{Nathan Isgur and Mark B. Wise, Phys. Lett. D 43 (1991) 819}
[{\WISBAR},{\WISBJ}]. The Bjorken sum rule recalls similar sum rules derived
for the
M\"ossbauer effect\REFS{\HJLMoss}{Harry J. Lipkin, Ann. Phys. 9 (1960) 332}
\REFSCON{\HJLSumR}{Harry J. Lipkin, Ann. Phys. 18 (1962) 182}
\REFSCON{\LipQM}{For a general review see Harry J.
Lipkin, Quantum Mechanics, North-Holland Publishing Co. Amsterdam
(1973) pp.33-110}[{\HJLMoss}-{\LipQM}]which can also be viewed as the
emission of an electroweak boson in a pointlike vertex by a heavy object
bound to a complicated system, where the transition involves a change in
the momentum and the mass of the heavy object without a change in the
interactions between the heavy object and the rest of the system. We
derive sum rules for the moments of the energy spectrum of the emitted
boson for decays of bound heavy quarks,
analogous to those previously derived for the M\"ossbauer effect
\REF{\PMOSSB}{For a general introduction see Harry J.
Lipkin, Argonne preprint ANL-HEP-PR-92-86
Submitted to Nuclear Physics A}[{\PMOSSB}].
In addition to the Bjorken sum rule for the decay rate we
find two additional sum rules showing that the mean and mean square
energies of the emitted boson or lepton pair are
independent of the detailed dynamics of the bound state, depend
only on the momentum spectrum of the heavy quark in the initial state
and are thus identical to those for the decay of a free
heavy quark in a gas with the same momentum spectrum as in the bound
state.

The weak decay from an initial state denoted by $i$ to a final state
denoted by $f$ is described by the Fermi
golden rule of time dependent perturbation theory
$$ W_{i\rightarrow f} = |\bra{f}H_{weak}
\ket{i}|^2 \rho(E_f)  \eqno(1a)
  $$
where $H_{weak}$ denotes the weak Hamiltonian and $\rho(E_f)$ is the
density of final states. For the case of a semileptonic decay where a lepton
pair is emitted with momentum $-\vec q$, the  matrix element for a transition
from an initial hadron state at rest $\ket{i_b}$ containing a $b$ quark to a
final state $\ket{f_c(\vec q) L(-\vec q)}$ of a hadron $f_c$ with total
momentum $\vec q$  containing a $c$ quark and a lepton pair with momentum
$-\vec q$ factorizes into a leptonic factor depending only on the lepton
variables and a weak vertex describing the heavy quark transition,
$$ \bra{f_c(\vec q) L(-\vec q)}H_{weak}\ket{i_b}=
g_k(\vec q)\bra{f_c}J_k(\vec q)\ket{i_b} \eqno(1b)  $$
where $g_k(\vec q)$ is a function of the lepton variables and $J_k(\vec q)$ is
the fourier component of the flavor-changing weak current for momentum
transfer $\vec q$ and the index k=0,1 describes the spin character of the
heavy quark transition; i.e. $J_1$ transforms like a vector under spin
rotations and describes spin-flip transitions while $J_o$ is simply the
identity operator and describes nonflip transitions.

The result (1) is exact to first order in the electroweak
interaction described by the standard model and exact to all orders in strong
interactions if the initial and final states considered $\ket{i_b}$ and
$\ket{f_c}$ are exact eigenstates of the strong interaction hamiltonian. We do
not need to assume the validity of QCD, perturbative or otherwise, or even the
validity of local field theory at this stage. All we need is the existence of
a hamiltonian for strong interactions and its eigenstates.

In heavy quark transitions this description applies to semileptonic decays
and also to nonleptonic decays in cases where factorization holds; i.e. there
are no final state interactions between the hadrons produced by the $W$ and the
final bound state.

We now derive the sum rules explicitly for the case of a heavy quark
transition (1). The relevant matrix elements of the weak current
$\bra{f_c}J_k(\vec q)\ket{i_b}$ depend only on the variables of the heavy quark
and determine completely the dependence of the transition on the hadronic wave
functions.

It is convenient to normalize the current and the coefficient $g_k(\vec q)$
$$ g_k(\vec q)\bra{f_c}J_k(\vec q)\ket{i_b} = g_k(q)
\bra{f_c}\Sigma_k e^{i\vec q \cdot \vec X}\ket{i_b}  \eqno (2a) $$
where the factor $g_k(q)$ depends upon the strength of the
interaction and depends upon the kinematic variables only via the magnitude of
the momentum transfer $\vec q$ and not upon the initial momentum of the heavy
quark and $\Sigma_k$ is a spin factor acting on the spin of the heavy quark and
normalized to satisfy the relation
$$ \sum_{\ket {f_c}} |\bra{f_c}\Sigma_k \ket{i_b} |^2 = 1  \eqno (2b) $$
The transition probability or branching ratio is
proportional to the square of this matrix element and multiplied by kinematic;
e.g. phase space factors which do not depend on the explicit form of the
hadron wave functions but only on hadron masses and the momenta of the external
particles. This matrix element is seen to satisfy the sum rule
$$ \sum_{\ket {f_c}} |g_k(\vec q)\bra{f_c}J_k(\vec q)\ket{i_b} |^2 = |g_k(q)|^2
 \eqno (2c) $$
This is just the Bjorken sum rule. To obtain additional sum rules
it is convenient to define the reduced matrix element
$$ \bra{f_c}M_k(\vec q)\ket{i_b} =
\bra{f_c}\Sigma_k e^{i\vec q \cdot \vec X}\ket{i_b}  \eqno (3a) $$
which expresses the Bjorken sum rule in the form
$$ \sum_{\ket {f_c}} |\bra{f_c}M_k(\vec q)\ket{i_b}|^2 =1 \eqno (3b) $$

The Hamiltonian of the bound state of a heavy quark of
flavor $f$ and light quark ``brown muck" can be written
as a function of the co-ordinate $\vec X$, the momentum
$\vec P$ and the mass $m_f$ of the heavy quark, and
all the degrees of freedom in the brown muck denoted by $\xi_\nu$,
$$ H_f = H(\vec P, m_f, \vec X, \xi_\nu)
       \eqno(4)  $$
where all the flavor dependence is in the heavy quark mass $m_f$.
The transition operator (3a) contains a momentum displacement operator,
$$ e^{-i\vec q \cdot \vec X} \vec P e^{i\vec q \cdot \vec X}
= \vec P + \vec q \eqno(5a) $$
$$ e^{-i\vec q \cdot \vec X} H_f e^{i\vec q \cdot \vec X}
= H_f [(\vec P + \vec q), m_f, \vec X, \xi_\nu]    \eqno(5b) $$

We now derive sum rules for the moments of the energy
distribution $E_c$ of the charmed final state,
$$ \langle {[E_c(\vec q)]^n}\rangle_k
\equiv \sum_{\ket {f_c}} (E_c)^n |\bra{f_c}M_k(\vec q)\ket{i_b}|^2
 = \bra{i_b}\Sigma_k
( e^{-i\vec q \cdot \vec X} H_c e^{i\vec q \cdot \vec X})^n\Sigma_k
\ket{i_b} = $$
$$  = \bra{i_b}\Sigma_k
\{H [(\vec P + \vec q), m_c, \vec X, \xi_\nu] \}^n  \Sigma_k
\ket{i_b}  \eqno (6a) $$
$$ \langle {[E_c(\vec q)]^n}\rangle_k
= \bra{B_k} \{ H_b +
H [(\vec P + \vec q), m_c, \vec X, \xi_\nu]  -
H [(\vec P ), m_b, \vec X, \xi_\nu]
\}^n  \ket{B_k}  \eqno (6b) $$
$$ \langle {[E_c(\vec q)]^n}\rangle_k
= \bra{B_k} \{ H_c +
H [(\vec P + \vec q), m_c, \vec X, \xi_\nu]  -
H [(\vec P ), m_c, \vec X, \xi_\nu]
\}^n  \ket{B_k}
\eqno (6c) $$
where $\ket{B_k}$ denotes the state produced by the operator
$\Sigma_k$ acting on the initial state $i_b$.
$$
\ket {B_k} \equiv \Sigma_k \ket{i_b} \eqno (7) $$
We have thus obtained sum rules for moments of the final state energy
distribution at a fixed momentum transfer $\vec q$ in terms of expectation
values calculated in the initial state $\ket{i_b}$.

The case $n=0$ is just the Bjorken sum rule, showing that the sum of the
squares of all transition matrix elements at fixed momentum transfer
$\vec q$ is completely independent of the hamiltonian $H$ describing the
dynamics of the bound state and is the same as for a free $b$ quark
decaying to charm.

The sum rules reduce to a particularly simple form for the first and
second moments, n=1 and n=2, if one simple additional assumption is made
which holds in all conventional models; namely that the state
$B_k$, produced by the operator $\Sigma_k$ acting on the initial state
$i_b$ is an eigenfuntion of the bound state Hamiltonian $H_b$
$$
H_b \ket {B_k} \equiv H_b \Sigma_k \ket{i_b} = M(B_k) \ket {B_k}
\eqno (8) $$
This assumption holds trivially for the case $k=0$ where the spin
operator $\Sigma_o$ is just the identity. It is a good approximation for
the case $k=1$ if spin effects are neglected as well as in all models
where flipping the spin of the heavy $b$ quark in the $B$ meson
produces the vector $B^*$ state. These include both simple
constituent quark models as well as more general models which assume
the approximation of heavy quark symmetry.
It also holds in all simple models for the $\Lambda_b$ baryon where
the $b$ quark carries the spin of the baryon, all other degrees of
freedom are coupled to zero angular momentum and flipping the spin of
the $b$ quark simply flips the spin of the whole baryon.

For the first and second moments, n=1 and n=2, the operator $H_b$
appears in eq. (6b) only as acting either to the right on
$\ket{B_k}$ or to the left on $\bra{B_k}$ to give the mass eigenvalue
$M(B_k)$. Thus for n=1 and 2,
$$ \langle {[E_c(\vec q)]^n}\rangle_k=
\bra{B_k}\{M(B_k) + R(\vec q) + I_{bc} - \delta m
\}^n  \ket{B_k}  \eqno (9a) $$
where
$$
\delta m = m_b - m_c
\eqno (9b) $$
$$
R(\vec q) =
H [(\vec P + \vec q), m_c, \vec X, \xi_\nu]  -
H [(\vec P ), m_c, \vec X, \xi_\nu] = $$ $$ =
\sqrt {(\vec P + \vec q)^2 + m_c^2} - \sqrt {\vec P^2 + m_c^2}
\approx {{q^2} \over {2 m_c}}
\eqno (9c) $$
$$
I_{bc} = \delta m +
H [\vec P , m_c, \vec X, \xi_\nu]  -
H [\vec P , m_b, \vec X, \xi_\nu] \approx \vec P^2 \cdot
{{\delta m} \over {2 m_c m_b}}
\eqno (9d) $$
where the approximate equalities hold for the case where
the nonrelativistic approximation is used
for the heavy quark kinetic energy, which is assumed to contain all
the dependence of $H$ on $\vec P$ and the heavy quark mass.
The quantity $R(\vec q)$ is just the free recoil energy; i.e. the change
in kinetic energy of a free charmed quark when its momentum changes
from $\vec P$ to $\vec P + \vec q$.
The quantity $I_{bc}$ is just the ``isomer" or ``isotope" shift; i.e.
the change in the binding energy of the bound state when the mass of the
heavy quark changes from $m_b$ to $m_c$.

The two sum rules are expressed in terms of
expectation values of operators depending
only on the change in the Hamiltonian when the heavy quark momentum
$\vec P$ is replaced by $\vec P + \vec q$
and the heavy quark mass is changed from $m_b$ to $m_c$.
They thus depend only on the momentum distribution of the
heavy quark in the initial state and are independent of the brown muck
in all models where $\vec P$
enters only into the kinetic energy term in the bound state Hamiltonian
and the change in hyperfine energy with quark mass is neglected.
They are therefore the same as for the decay of free
$b$ quarks in a gas with the same momentum distribution as in the bound
state.

The sum rules can be written in various ways appropriate for different
applications. These depend upon which experimental quantities are
measured and upon assumptions about values of parameters which are not
directly measured like quark masses and hyperfine energy contributions.

The sum rules can be expressed in terms of
the energy $ E_W(\vec q) $ carried off by the $W$ and observed as the
energy of the lepton pair or hadrons produced from the $W$ which must
have momentum $\vec q$ for momentum conservation. From energy
conservation
$$ E_W(\vec q) = M(i_b) - E_c(\vec q)
\eqno (10)$$
Thus
$$ \langle {E_W(\vec q)}\rangle_k \equiv
\sum_{\ket {f_c}} E_W
|\bra{f_c}M_k(\vec q)\ket{i_b}|^2
= M(i_b) - \langle {[E_c(\vec q)]}\rangle_k
= $$
$$ = \delta m - \bra{B_k}R(\vec q) + I_{bc}\ket{B_k} + M(i_b) - M(B_k)
$$ $$ \approx
\delta m - {{q^2} \over {2 m_c}} -
\bra{B_k}\vec P^2\ket{B_k}\cdot {{\delta m} \over {2 m_c m_b}}
+ M(i_b) - M(B_k)
\eqno (11a) $$
$$ \langle {[E_W(\vec q)]^2}\rangle_k \equiv
\sum_{\ket {f_c}}[ E_W]^2
|\bra{f_c}M_k(\vec q)\ket{i_b}|^2 = $$
$$ =
\langle {[E_W(\vec q)]}\rangle_k^2  +
\bra{B_k}\{R(\vec q) + I_{bc}\}^2\ket{B_k} -
\bra{B_k}\{R(\vec q) + I_{bc}\}\ket{B_k}^2
$$ $$ \approx
\langle {[E_W(\vec q)]}\rangle_k^2 +
{{\bra{B_k} \vec P^2 \ket{B_k} \cdot q^2 }\over{3m_c^2}} +
{{(\delta m)^2} \over {4 m_c^2 m_b^2}} \cdot
(\bra{B_k} P^4\ket{B_k} - \bra{B_k} P^2\ket{B_k}^2)
 \eqno (11b) $$
The mean energy carried by the $W$ is seen to differ from the heavy
quark mass difference $\delta m$ by three corrections: the free
recoil kinetic energy, the ``isomer shift" and a spin correction
$ M(i_b) - M(B_k)$
which is present in the case of spin-flip transitions and is
equal to the hyperfine splitting.
If we make the nonrelativistic approximation for the heavy quark motion,
and neglect the dependence of the hyperfine energy on the heavy quark
mass the two sum rules reduce to simple expressions for the
mean energy and the dispersion of the energy
distribution of the emitted $W$ at fixed momentum $q$ when appropriate
kinematic factors (i.e. phase space)
are taken out of the observed distributions.

When the momentum transfer $\vec q$ and the mean momentum $\vec P$
of the heavy quark in the bound state are both small in comparison with the
heavy quark mass, the dispersion vanishes and the mean lepton energy is just
the quark mass difference $m_b - m_c$. The transition is always to the ground
state of the charmed system and the differences in the brown muck wave
functions
in the two cases is negligible.
When the momentum transfer $\vec q$ is negligibly small, the
correction resulting from the finite momentum of the heavy quark in the
bound state is just the difference in the heavy quark kinetic energies in the
initial and final ground states. This is just equal to the difference in
binding energies to first order in the mass difference.

We now derive a sum rule for the mean excitation energy of
the charmed final state above the ground state of the charmed system. The
the exact ground state energy is not known since it
depends upon the binding energy of the brown muck to the heavy quark.
We express this ignorance by writing
$$  \bra{B_k} H_c \ket{B_k} = M(D_k) +
\epsilon     \eqno(12)   $$
where $M(D_k)$ is the mass of the charmed analogue of the state $B_k$.
The LHS of eq. (12) is just the mass $M(D_k)$ in the heavy quark
symmetry limit and is exact to first order in the difference
$({{1}\over{m_c}}-{{1}\over{m_b}})$. The correction is therefore second
order and can also be seen to be positive, since the perturbation
result is also variational and gives an upper bound for the mass,

The mean excitation energy $\Delta E_c$ of the observed
spectrum above the ground state is obtained by combining eqs.
(6c), (9c) and (12) to obtain
$$ \langle \Delta E_c \rangle_k \equiv
\langle E_c \rangle_k - M(D_k) \equiv
\sum_{\ket {f_c}}
E_c|\bra{f_c}M_k(\vec q)\ket{i_b}|^2 - M(D_k) =
\bra{B_k} R(\vec q) \ket{B_k} +
\epsilon     \eqno(13)   $$
In nearly all models generally considered all the dependence of $H$ on
the heavy quark momentum $P$ is in the kinetic energy; i.e. all
dependence of the forces between the heavy quark and the brown muck upon
the velocity of the heavy quark is neglected. In that case $R(\vec q)$
is just the free recoil energy; the kinetic energy gained by a
free quark with momentum $\vec P$ as a result of
absorbing a momentum transfer $\vec q$.
Thus the mean excitation energy including
corrections to the binding energies which are first order in the
difference of 1/m is just the free recoil energy $R(\vec q)$.

In contrast to the M\"ossbauer effect, where the whole system has a very
high mass and the recoil kinetic energy of the whole system is negligible,
the contribution of the recoil kinetic energy of the final charmed state
to the mean excitation energy (13) is appreciable. We therefore
express the sum rule (13) in terms of the invariant mass
$M_f$ of the final state $\ket{f_c}$ and the
excitation energy of the first $p$-wave resonance in the charmed system
denoted by $\Delta E_{1k}$.
$$ {{ \langle \sqrt {M_f^2 + \vec q^2 } \rangle_k  -
\sqrt {M(D_k)^2 + \vec q^2 }
}\over{\Delta E_{1k}}}
= {{
\bra{B_k} R(\vec q) \ket{B_k} - \sqrt {M(D_k)^2 + \vec q^2 } + M(D_k)
+ \epsilon}\over{\Delta E_{1k}}}
\eqno(14)   $$
The sum rule (14) can be simplified by introducing parameters
$$ \alpha \equiv{{\sqrt{M(D_k)^2 + \vec q^2 }}\over{M(D_k)}}
\eqno(15a)   $$
$$ \mu \equiv M(D_k) - \bra{D_k} \sqrt {\vec P^2 + m_c^2} \ket{D_k}
\leq M(D_k) - m_c
\eqno(15b)   $$
Then
$$ \sqrt {M_f^2 + \vec q^2 } -
\sqrt {M(D_k)^2 + \vec q^2}=
{{M_f^2 - M(D_k)^2
}\over{\sqrt {M_f^2 + \vec q^2 } +
\sqrt {M(D_k)^2 + \vec q^2 }}}
\geq {{M_f - M(D_k)}\over{\alpha}}
\eqno(16)   $$
$$ \vbox{\eqalignno{
\bra{B_k} R(\vec q) \ket{B_k}
- & (\alpha - 1) M(D_k)
=  \sqrt{\alpha^2 M(D_k)^2 -  2 M(D_k) \mu + \mu^2}
- \alpha M(D_k) + \mu
& ~ ~ ~ ~ ~ \cr
= & {{2(\alpha -1) M(D_k) \mu }\over{
\sqrt{\alpha^2 M(D_k)^2 -  2 M(D_k) \mu + \mu^2}
+ \alpha M(D_k) - \mu}}
&(17a)   \cr
\bra{B_k} R(\vec q) \ket{B_k}
- & (\alpha - 1) M(D_k)
\leq  {{(\alpha -1) M(D_k) \mu }\over{
\alpha M(D_k) - \mu}}
& (17b)  \cr}} $$
where we have assumed that $\vec P \cdot \vec q$ which
averages to zero over the angular distribution of $\vec P$ can be
neglected and that
the correction $\epsilon$ to the first order result for $M(D_k)$
is negligible in comparison with excitation energies,
$$ \alpha \epsilon  << \Delta E_{1k} \eqno(18)   $$
Substituting eqs. (15-18) into the sum rule(14) gives
$$
{{\langle M_f \rangle_k - M(D_k)}\over{
\Delta E_{1k}
}}\leq
{{\mu}\over{\Delta E_{1k}}}\cdot
{{\alpha (\alpha -1) M(D_k)}\over{
\alpha M(D_k) - \mu}}
\leq {{M(D_k) - m_c}\over{\Delta E_{1k}}}\cdot
{{\alpha (\alpha -1) M(D_k)
}\over{
(\alpha -1) M(D_k) + m_c}}
\eqno(19)   $$

The excitation spectrum of the final charmed states is seen to have
a mean excitation energy above the
ground state $D_k$ whose scale is set by the parameter $\mu$.
This parameter is just the difference between the
charmed hadron mass and the energy of the charmed quark and is roughly
equal to a light constituent quark mass and bounded by eq. (15b).
For low values of $q^2$
the mean excitation energy is even lower by the ratio $q^2/2
M(D_k)^2$.

{}From these inequalities (19) we obtain upper bounds for the probability
of B decays into excited charmed states.
$$ \sum_{\ket {f_c} \not= \ket{D_k}}
|\bra{f_c}M_k(\vec q)\ket{i_b}|^2 \leq
{{M(D_k) - m_c}\over{\Delta E_{1k}}}\cdot
{{\alpha (\alpha -1) M(D_k)
}\over{
(\alpha -1) M(D_k) + m_c}}
\eqno(20a)   $$
$$ \sum_{\ket {f_c} \not= \ket{D_k}}
|\bra{f_c}M_k(\vec q)\ket{i_b}|^2 \leq
{{M(D_k) - m_c}\over{\Delta E_{1k}}}\cdot
{{\alpha q^2
}\over{
q^2 +
(\alpha +1) M(D_k) m_c}}
\eqno(20b)   $$
while for any given excited final state other than $\ket{D_{k1}}$
$ \ket {f_c} \not= \ket{D_k}$,
$$ |\bra{f_c}M_k(\vec q)\ket{i_b}|^2 \leq
{{M(D_k) - m_c}
\over{M_c - M(D_k)}}\cdot
{{\alpha (\alpha -1) M(D_k)
}\over{
(\alpha -1) M(D_k) + m_c}}
\eqno(21)   $$
This result eqs. (20) is an upper bound for the total probability of
decays into states other than the ground state, when the phase space
factors are assumed to be equal for all final states. Including phase
space factors will give an even stronger upper bound.
The maximum value of $q^2$ which leaves the energy available just at the
threshold for producing the lowest p-wave charmed
state\REF{\PDG}{Particle Data Group, Phys. Lett. B239 (1990)
1}$[{\PDG}]$,
$D(2420)$ is
$q^2 \approx 1.25 M(D_k)^2$, which gives $\alpha = 1.5$. This occurs
when the lepton momenta are
exactly parallel and they carry the minimum possible energy for a given
value of $q^2$.
This gives an upper bound on the probability of producing an excited
charmed meson state of less than 56\% for the quoted
values$[{\PDG}]$, $M(D) = 1.87$ GeV, $m_c = 1.35$ GeV,
$\Delta E_{1k} = 0.55 $ GeV.
An even smaller upper bound is obtainable if some estimate of
the kinetic energy of the heavy quark can be used in eq. (15b).
An upper bound for the probability
of producing a given state $\ket{f_c}$ with higher mass is given by
eq. (21) for the extreme case where no lower excited states are
produced.
The probability of producing the ground state is thus
$\approx 50\%$ even under these extreme conditions and increases rapidly
with more realistic lower values of $q^2$.
Thus the excitation spectrum will be dominated by the ground state $D_k$
and the first low-lying $D^*$'s.

Stimulating and clarifying discussions with Nathan Isgur and Jonathan L.
Rosner are gratefully acknowledged. This research was partially supported
by the Basic Research Foundation administered by the Israel Academy of
Sciences and Humanities and by grant No. 90-00342 from the United
States-Israel Binational Science Foundation (BSF), Jerusalem, Israel.

\refout
\end